\documentstyle[psfig]{mn}

\newif\ifAMStwofonts

\def\lapp{\ifmmode\stackrel{<}{_{\sim}}\else$\stackrel{<}{_{\sim}}$\fi}
\def\gapp{\ifmmode\stackrel{>}{_{\sim}}\else$\stackrel{>}{_{\sim}}$\fi}

\title[Profile morphology and polarization of young pulsars]
{Profile morphology and polarization of young pulsars}
\author[Johnston \& Weisberg]
{Simon Johnston$^1$ and Joel M. Weisberg$^2$\\
$^1$Australia Telescope National Facility, CSIRO, P.O. Box 76, 
Epping, NSW 1710, Australia. \\
$^2$Dept. of Physics and Astronomy, Carleton College, Northfield, MN 55057, USA.
}
\date{\today}
\pagerange{\pageref{firstpage}--\pageref{lastpage}}
\pubyear{2006}
\begin{document}
\maketitle
\label{firstpage}

\begin{abstract}
We present polarization profiles at 1.4 and 3.1~GHz for 14 young pulsars 
with characteristic ages less than 75 kyr. Careful calibration ensures
that the absolute position angle of the linearly polarized radiation
at the pulsar is obtained. In combination with previously published
data we draw three main conclusions about the pulse profiles
of young pulsars. (1) Pulse profiles are simple and consist of either
one or two prominent components. (2) The linearly polarized fraction is nearly
always in excess of 70 per cent. (3) In profiles with two components
the trailing component nearly always dominates, only the trailing
component shows circular polarization and the position angle swing
is generally flat across the leading component and steep across
the trailing component.

Based on these results we can make the following 
generalisations about the emission beams of young pulsars. (1) There
is a single, relatively wide cone of emission from near the last
open field lines. (2) Core emission is absent or rather weak.
(3) The height of the emission is between 1 and 10 
per cent of the light cylinder radius.
%
\end{abstract}


\section{Introduction}
The polarization profiles of radio pulsars have long been a valuable
tool for a variety of different purposes. They can be used, for example,
to classify the different types of profile,
to understand the underlying emission 
mechanism and to determine the geometry of the star.
In the rotating vector model (RVM) of Radhakrishnan \& Cooke
(1969)\nocite{rc69a}, the radiation is beamed along the field lines
and the plane of polarization is determined by the angle of the
magnetic field as it sweeps past the line of sight.
The position angle (PA) as a
function of pulse longitude, $\phi$, can be expressed as
\begin{equation}
{\rm PA} = {\rm PA}_{0} +
{\rm arctan} \left( \frac{{\rm sin}\alpha
\, {\rm sin}(\phi - \phi_0)}{{\rm sin}\zeta
\, {\rm cos}\alpha - {\rm cos}\zeta
\, {\rm sin}\alpha \, {\rm cos}(\phi - \phi_0)} \right)
\end{equation}
Here, $\alpha$ is the angle between the rotation axis and the magnetic
axis and $\zeta=\alpha+\beta$ with $\beta$ being the angle of closest approach
of the line of sight to the magnetic axis.
$\phi_0$ is the pulse longitude at which the PA is
PA$_{0}$, which also corresponds to the PA of the rotation axis 
projected onto the plane of the sky. We note that RVM fitting does not
depend on the total intensity profile, or this location of the profile
symmetry points.
Unfortunately, for most pulsars it is difficult to
determine $\alpha$ and $\beta$ with any degree of accuracy, partly
because the longitude over which pulsars emit is rather small and
partly because strong deviations from a simple swing of PA are often
observed.  This makes the determination of the various angles
straightforward only in the $\sim$15 per cent of pulsars for 
which the RVM works (see the discussion in 
Everett \& Weisberg 2001\nocite{ew01}).

If the emission occurs at some height above the pulsar surface,
the PA swing can be delayed with respect to the total intensity 
profile by relativistic effects such as aberration and
retardation (hereafter referred to as A/R) as initially
discussed by Blaskiewicz,
Cordes \& Wasserman (1991)\nocite{bcw91}. The magnitude of the shift
in longitude, $\delta\phi$(PA) (in radians),
is related to the emission height relative to the centre
of the star, $h_{\rm em}$, via
\begin{equation}
\delta\phi({\rm PA}) = \frac{8\,\,\, \pi\,\,\, h_{\rm em}}{P\,\,\, c}
\end{equation}
where $P$ is the pulsar period and $c$ the speed of light
(Blaskiewicz et al. 1991)\nocite{bcw91}.
We note that, at least to first order, the PA swing is not altered by
A/R effects other than the longitude delay. These effects therefore
do not affect the measured values of $\alpha$ and $\beta$ from equation 1.

There is also a geometrical method which can be used to compute
emission heights if the angles $\alpha$ and $\beta$ are known.
Under the assumption of a dipolar field and a circular emission zone,
the half-opening angle of the emission cone, $\rho_e$ can be expressed as
\begin{equation}
{\rm cos}\rho_e = {\rm cos}\alpha\,\, {\rm cos}\zeta\,\, +\,\, {\rm sin}\alpha\,\, {\rm sin}\zeta\,\, {\rm cos}(W/2)
\end{equation}
where $W$ is the measured pulse width in longitude \cite{ggr84}.
If one assumes that the emission
extends to the final open field line then the emission height can
be derived through the expression
\begin{equation}
\rho_o = 3\,\,\, \sqrt{\frac{\pi \,\,\, h_{\rm em}}{2\,\,\, P\,\,\, c}}
\end{equation}
\cite{ran90}. Here we use $\rho_o$ to denote the half-opening angle
of the cone at the last open field line, and note that unless the
emission extends right to the edge of the cone then
$\rho_o > \rho_e$.

It is possible to compute relative emission heights of the radiation without
the polarization information \cite{gg01}.  Imagine the cone emission comes
from a circular symmetric region around the core. Then, if cone emission
arises higher in the magnetosphere than the core emission, the entire
circular structure will be shifted relative to the core due to A/R
effects and the profile will appear asymmetric.
If a given profile contains well defined core and conal emission
then the difference in separation between the leading and trailing
conal components and the core, $\delta\phi$(CC), can be used to compute 
their relative emission heights, $\Delta h_{\rm em}$.
In this case
\begin{equation}
\delta\phi({\rm CC}) = \frac{4 \,\,\,\pi \,\,\,\ \Delta h_{\rm em}} {P\,\,\,c}
\end{equation}
\cite{drh04}. Gangadhara (2005)\nocite{gan05} has modified this 
equation to take into
account the viewing geometry which can affect the derived emission heights
by $\sim$10 per cent in some cases.
It can be useful to compare the emission height with the radius of
the light cylinder, $r_{\rm lc} = Pc/2\pi$. Generally it is found
that the radio emission occurs in regions significantly below about
0.1~$r_{\rm lc}$.
Substantial discussion of the merits and failings of all these methods
can be found in recent papers by e.g.
Gupta \& Gangadhara (2003)\nocite{gg03}, Mitra \& Li (2004)\nocite{ml04},
Dyks et al. (2004)\nocite{drh04} and Gangadhara (2005)\nocite{gan05}.

Young pulsars and/or pulsars with high spin down energies,
as a group, tend to have high ($>70$ per cent)
linear polarization (e.g. Qiao et al. 1995\nocite{qmlg95},
Crawford et al. 2001\nocite{cmk01})
and often show simple Gaussian-like profiles with little structure.
Some however, such as PSRs B1259$-$63 \cite{mj95} and B0906$-$49
\cite{qmlg95} have wide double profiles.
Also, as a class, young pulsars tend to have flatter spectral indices
than the older population. Many have associated non-thermal high energy
emission whose profiles are offset from the radio profiles.
These features led Manchester (1996)\nocite{man96} to suggest that emission
from young pulsars occurred relatively near the light cylinder.

Han et al. (1998)\nocite{hmxq98} showed a remarkable correlation
between the circular
polarization properties of pulsars and the direction of their position
angle swing. For pulsars which show simple (conal) double profiles, the
sign of circular polarization under the components is correlated with
the slope of the PA swing; left-hand circular polarization implies
a negative slope and vice-versa. There appear to be no exceptions to
this rule in a sample which has now grown to 40 pulsars \cite{yh05}.
However, this correlation does not hold for other pulsars
with more complicated pulse profiles.

The supernova explosion that creates the pulsar also produces a supernova
remnant (SNR) and many young pulsars are seen, at least in projection,
inside SNR shells. One way to confirm the association between the pulsar
and the SNR is to obtain the proper motion for the pulsar and determine
whether it originated from the SNR centre.
Only a relatively small number of young pulsars have
good proper motion measurements but polarization observations also
provide valuable information.
Johnston et al. (2005) have recently used the polarization from young
pulsars to show that, for the majority of cases, the rotation axis 
is aligned with the velocity vector.
Therefore, for pulsars without known proper motions, it may be possible
to obtain the direction of motion directly from the direction of the
rotation axis inferred through polarization measurements. 

The known pulsars can be sorted according to their characteristic ages
($\tau_c = P/2\dot{P}$, where $P$ is the pulsar spin period and $\dot{P}$
is the period derivative) or their spin down energy, $\dot{E}$
(proportional to $\dot{P}/P^3$). Naturally, the youngest pulsars tend
also to have the highest $\dot{E}$. One of the aims of this paper is
to determine whether, as a class, the young, highly energetic pulsars
have characteristic total intensity and polarization profiles.

A significant number of young pulsars already have measured polarization
properties by a variety of authors over a range of frequencies albeit
with rather low time resolution and without absolute PA determination.
However, recent pulsar surveys have uncovered many more young objects
for which the total intensity profiles are available but with few
polarization measurements. We selected 10 of these pulsars to observe
at both 1.4 and 3.1~GHz, and a further 4 which were observed at 1.4~GHz
only. The selection was based on the following
criteria: (1) Characteristic ages of less than 75 kyr,
(2) Right Ascension less than 17~h, (2) Declinations less
than 0\degr. Many of the pulsars also have a possible association with 
high energy emission or with an SNR.

The outline of the paper is as follows. In Section~2 we describe in more
detail those pulsars which have associations with high energy emission
and/or SNRs.
In Section~3 we describe the observations and present the
results in Section~4, with Section~5 detailing fits of the
rotating vector model.
In Section~6 we discuss
the polarization of young pulsars generally, and 
in Section~7 we analyse the implications of
the polarization results for the proper motion direction.

\section{Pulsars with possible high energy or SNR associations}
PSR~J1015$-$5719 was discovered by Kramer et al. (2003)\nocite{kbm+03} and
they and Torres et al. (2001)\nocite{tbc01} pointed out that it is located near
the unidentified EGRET source 3EG J1014$-$5705. The energetics 
make the association plausible if the distance to the pulsar
is $\sim$5~kpc.

PSR~J1016$-$5857 was discovered by Camilo et al. (2001)\nocite{cbm+01}.
It is located
in projection near SNR~G284.3$-$1.8 and although the distance to
the SNR and to the pulsar disagree, Camilo et al. (2001) consider
the association to be likely. The pulsar is located about 15 arcmin
from the apparent SNR centre in a WNW direction. If the association
is correct, the proper motion direction should be $\sim$300\degr.
The pulsar may be
associated with the EGRET source 3EG~J1013$-$5915.

PSR~J1105$-$6107 was discovered by Kaspi et al. (1997)\nocite{kbm+97}.
It is a young pulsar and is not obviously embedded in a parent SNR. However, 
SNR~G290.1$-$0.8 is nearby (in projection) and Kaspi et al. (1997)
discuss the possibility that the two are associated. If so, its proper
motion should be in a SE direction (135\degr) but this has not 
yet been measured.  There is a possible association with the EGRET source
2EG~J1103$-$6106. Crawford et al. (2001)\nocite{cmk01} provide a polarization
profile of this pulsar at 1.4~GHz and computed its RM to be 166~rad~m$^{-2}$.

PSR~J1119$-$6127 was discovered by Camilo et al. (2000)\nocite{ckl+00}
and its environs have been imaged in the
radio (Crawford et al. 2001)\nocite{cgk+01} and the
X-ray (Pivovaroff et al. 2001)\nocite{pkc+01}.
It lies virtually in the centre of the shell-like
SNR~G292.2$-$0.5 and the association appears secure.
Gonzalez et al. (2005)\nocite{gkm+05} have recently detected a pulsar
wind nebula and thermal pulsations from the pulsar in the X-ray.
Crawford \& Keim (2003)\nocite{ck03} published polarization results for 
this pulsar at 1.4~GHz and showed that it had a high degree of 
linear polarization and an RM of 842~rad~m$^{-2}$.

PSR~J1341$-$6220 (B1338$-$62) was discovered by 
Manchester et al. (1985)\nocite{mdt85}. Timing of
the pulsar proved it to be young (Kaspi et al. 1992)\nocite{kmj+92}
and a high resolution image by Caswell et al. (1992)\nocite{cks+92}
revealed the pulsar to be located near the centre of SNR~G308.8$-$0.1.
One might then expect the proper motion direction to be $\sim$315\degr.
The pulsar/SNR system is rather similar
to the PSR~B1509$-$58/SNR G320.4$-$1.2 association discussed in more
detail below. Qiao et al. (1995)\nocite{qmlg95} and
Crawford et al. (2001)\nocite{cmk01} provide polarization profiles
of this pulsar at 1.4~GHz which show a profile largely dominated
by scattering. The RM of the pulsar is $-$946~rad~m$^{-2}$.

PSRs J1412$-$6145 \cite{mlc+01} and J1413$-$6141 \cite{kbm+03}
were discovered as part
of the Parkes multibeam survey. Both pulsars lie within the SNR G312.4$-$0.4
though it is not clear which, if either, is associated with it
(Doherty et al. 2003)\nocite{djg+03}. Proper motions for these pulsar would help
resolve this issue. The error box of the EGRET source 3EG~J1410$-$6147 
encompasses the SNR and the two pulsars.
Both Case \& Bhattacharya (1999)\nocite{cb99}
and Doherty et al. (2003)\nocite{djg+03} discuss the possibility that 
the EGRET source and the SNR and/or the pulsar(s) are associated without
reaching any firm conclusions.

PSRs J1420$-$6048 was discovered by D'Amico et al. (2001)\nocite{dkm+01}.
Prior to the detection of the pulsar,  Roberts et al. (1999)\nocite{rrjg99}
made a radio image of the area surrounding the EGRET source 2EG J1418$-$6049
and detected a wind nebula (the `Rabbit') surrounded by a possible
non-thermal shell (the `Kookaburra'). They concluded that a young
energetic pulsar was likely
the source of both the wind nebula and the EGRET source.
However, the discovered pulsar lies somewhat outside the wind nebula,
further muddying 
an already complex picture.  X-ray pulsations have been detected from 
the pulsar (Roberts et al. 2001\nocite{rrj01})
and it remains a candidate for the EGRET source.
Roberts et al. (2001) also present the polarization profile of the
pulsar at 1.4~GHz and derive an RM of $-$106~rad~m$^{-2}$.

PSR J1513$-$5908 (B1509$-$58) was first discovered in 
X-rays (Seward et al. 1982)\nocite{sh82}
and subsequently in the radio (Manchester et al. 1982)\nocite{mtd82}
and is likely associated and interacting with SNR G320.4$-$1.2.
In an exhaustive radio study of the complex,
Gaensler et al. (1999)\nocite{gbm+99} make two comments of
relevance here. First they speculate that the star Muzzio 10 may have been
a (former) companion of the pulsar and hence the pulsar's proper motion
should be at a position angle of 168\degr. From morphology considerations
they argue that the pulsar's rotation axis must point at an angle of
approximately 145\degr\ and 315\degr\
(see also Brazier \& Becker 1997\nocite{bb97}).
Observations of the wind nebula around the pulsar in the X-ray, showed
that the nebula has a clear symmetry axis with position angle of 150\degr\
\cite{gak+02}, consistent with the radio observations.
Polarization profiles for this
pulsar were shown at 0.66 and 1.4~GHz by Crawford et al. (2001). At
0.66~GHz the pulse is extremely scatter broadened but at 1.4~GHz shows
a simple Gaussian profile extending over about 70\degr\ of longitude.

\section{Observations}
Observations were carried out using the 64-m radio telescope located
near Parkes, New South Wales. Two different receiver packages were used
at frequencies near 1.4 and 3.1~GHz. The H-OH receiver covers the
frequency range from 1.2 to 1.7 GHz. We used a central frequency of
1.368~GHz and a total bandwidth of 256~MHz. The 10cm part of the
dual frequency 10/50~cm receiver has a total bandwidth of 1024~MHz
centered at 3.1~GHz. All the observations took place in 2005.
The majority of the 1.4~GHz observations were made in the period
July 9 to 13 and the 3.1~GHz observations made between June 20 and 26.

The pulsars were typically observed for 30~mins each, preceded by a 2~min
observation of the pulsed calibrator. The total bandwidth was subdivided
into 1024 frequency channels and the pulsar period divided into 1024
phase bins by the backend correlator. The correlator folds the data
for 60~s at the topocentric period of the pulsar at that epoch and
records the data to disk for offline-processing.
During the observing session, observations
were made of the flux calibrator Hydra~A whose flux density is 43 and
21~Jy at the observing frequencies of 1.4 and 3.1~GHz.
This allowed us to determine the system equivalent
flux density to be 28.7 and 46.2~Jy and allowed for flux calibration of the
pulsar profiles.

The data were analysed off-line using the PSRCHIVE software package
(Hotan et al. 2005). Polarization calibration was carried out using
the observations of the pulsed calibrator signal to determine the relative
gain and phase between the two feed probes. The data were corrected
for parallactic angle and the orientation of the feed. The position
angles were also corrected for the Faraday rotation along the path to the
pulsar (i.e. the ionospheric and interstellar medium contributions)
using the rotation measure (RM) determined from the data.
Hence, all PAs are absolute PAs at the pulsar and can thus be compared
directly at different frequencies.
Flux calibration was carried out using the Hydra~A observations.
The final product was therefore flux and polarization calibrated
Stokes $I$, $Q$, $U$, $V$ profiles.
\begin{table*}
\caption{Pulsar parameters}
\begin{tabular}{llrrllrrrrr}
\hline & \vspace{-3mm} \\
Jname & Bname & Period & Age & T$_{1.4}$/T$_{3.1}$ & N$_{1.4}$/N$_{3.1}$ &
\multicolumn{1}{c}{RM} & S$_{1.4}$ & S$_{3.1}$ & W10$_{1.4}$ & W10$_{3.1}$ \\
& & (ms) & (kyr) & \multicolumn{1}{c}{(s)} & & rad~m$^{-2}$ & (mJy) &
(mJy) & (deg) & (deg) \\
\hline & \vspace{-3mm} \\
J0729$-$1448 &            & 251.7 & 35.2 & 5400/6000 & 256/1024 & 46$\pm$1 & 0.50 & 0.43 & 25 & 24 \\
J0940$-$5048 &            &  87.5 & 42.2 & 3600/6000 & 256/512 & $-$31$\pm$1 & 0.65 & 0.47 & 43 & 42 \\
J1015$-$5719 &            & 139.9 & 38.6 & 1800      & 256  & 96$\pm$2 & 3.50 &      & 155  &   \\
J1016$-$5857 &            & 107.4 & 21.0 & 3600/6000 & 256/512 & $-$540$\pm$3 & 0.81 & 0.38 & 36 & 31 \\
J1105$-$6107 &            &  63.2 & 63.6 & 1800/3600 & 256/512 & 187$\pm$1 & 1.25 & 0.63 & 23 & 25 \\
J1119$-$6127 &            & 407.8 &  1.6 & 3600/3600 & 256/256 & 853$\pm$2 & 0.99 & 0.44 & 43 & 34 \\
J1301$-$6305 &            & 184.5 & 11.0 & 1800      & 256  & $-$625$\pm$25 &  1.00 &  & 80   &    \\
J1341$-$6220 & B1338$-$62 & 193.3 & 12.1 & 1800/1200 & 256/512 & $-$921$\pm$3 & 2.75 &      & 42 & 15 \\
J1357$-$6429 &            & 166.1 &  7.3 & 1800      & 256  & $-$47$\pm$2 &      &   & 45 &    \\
J1412$-$6145 &            & 315.2 & 50.6 & 3600/1200 & 256/256 & $-$130$\pm$13 & 0.64 & 0.24 & 21 & 18 \\
J1413$-$6141 &            & 285.6 & 13.6 & 5400/2400 & 256/256 & $-$400$\pm$40 & 0.85 & 0.54 & 72 & 7 \\
J1420$-$6048 &            &  68.2 & 13.0 & 5400/2400 & 256/512 & $-$122$\pm$5 & 1.20 &      & 73 & 58 \\
J1513$-$5908 & B1509$-$58 & 150.7 &  1.6 & 5400/5400 & 256/128 & 216$\pm$1 & 1.66 & 0.46 & 103 & 127 \\
J1702$-$4310 &            & 240.5 & 17.0 & 5400      & 256     & $-$35$\pm$12 & 1.12 & & 35 \\
\hline & \vspace{-3mm} \\
\end{tabular}
\end{table*}

The contributions to the
measured values of RM arising in the ionosphere have been estimated
by integrating the International Reference Ionosphere and
International Geomagnetic Reference Field time-dependent models
of the ionospheric electron
density and geomagnetic field through the ionosphere along the sight
line between the telescope and the pulsar \cite{bil03} using 
an implementation provided by JL Han.
Values of the ionospheric
RM range from 0 to $-$2~rad~m$^{-2}$ and
were subtracted from the measured values
to leave only the interstellar components.

\section{Results}
Table~1 lists the pulsars observed along with their period (column 3) and
characteristic age (column 4). Column 5 displays the total
integration time at 1.4 and 3.1~GHz and column 6 gives the number of phase
bins in the profiles at the two frequencies. Column 7 gives the
RM due to the interstellar medium alone, columns 8 and 9 list the 
flux densities at the two observing
frequencies. The final two columns give the width of the profile at the
10 per cent intensity level.

Figures 1-6 show the calibrated polarization profiles for the pulsars
in our sample. We have located zero longitude at the symmetrical midpoint
given by the 10 per cent intensity level of the profiles in most cases.
The exceptions to this rule are the 1.4~GHz profiles of PSRs~J1341$-$6220
and J1413$-$5141 which are scatter broadened; the peak of the profile was
aligned with that at 3.1~GHz.
Results for individual pulsars are described in detail below.

\noindent
{\bf PSR J0729$-$1448:} The profile of this pulsar undergoes significant
frequency evolution between 1.4 and 3.1~GHz. At both frequencies there
are 3 clearly visible components; this is one of only two pulsars in our
sample to show a central component. At the lower frequency, the trailing
component dominates and the leading component is weakest, whereas at
the higher frequency the leading and trailing components have roughly
equal strength. The linear polarization at both frequencies 
is high throughout with positive circular polarization seen against
the trailing component. The PA swing steepens sharply through the trailing 
component. In spite of the strong frequency evolution, the profiles widths
are virtually identical at both frequencies.

\noindent
{\bf PSR J0940$-$5428:} The pulse profile is very similar at both 1.4
and 3.1~GHz showing what appears to be a classic double structure with
a stronger trailing component. The saddle region is somewhat better filled
in at the higher frequency even though the overall pulse widths
are similar at both frequencies. The linear polarization is high at 
both frequencies and there is a hint of circular polarization against
the trailing component. The PA swing is steeper against the second
component.
\begin{figure*}
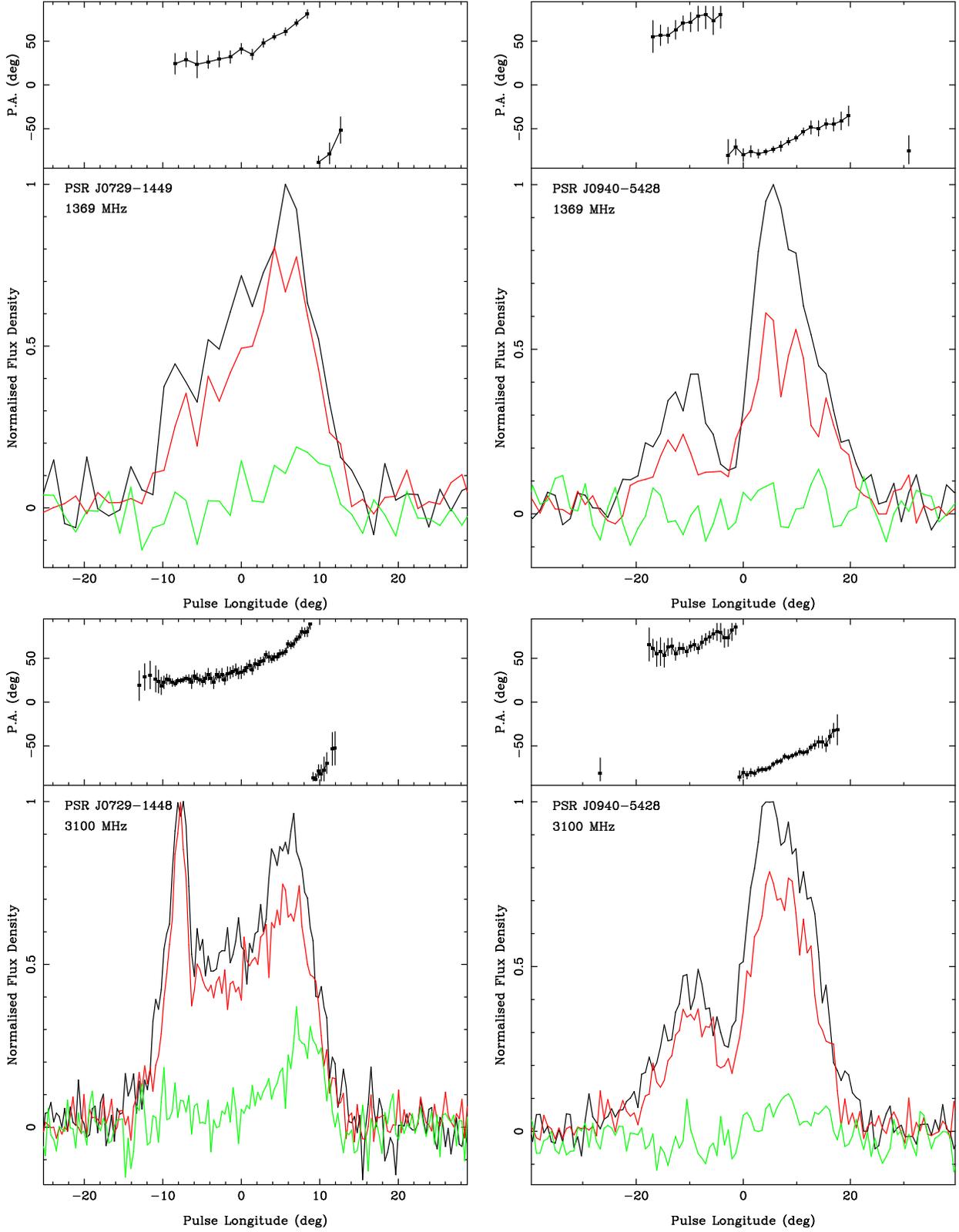

\begin{tabular}{cc}
\psfig{figure=fig1a.ps,angle=0,width=8cm} &
\psfig{figure=fig1b.ps,angle=0,width=8cm} \\
\psfig{figure=fig1c.ps,angle=0,width=8cm} &
\psfig{figure=fig1d.ps,angle=0,width=8cm} \\
\end{tabular}
\caption{Polarization profiles at 1.4~GHz and 3.1~GHz for
PSRs J0729$-$1449 and J0940$-$5428. The top panel of each plot shows the 
PA variation with respect to celestial north as a function of longitude.
The PAs are corrected for RM and
represent the (frequency independent) value at the pulsar.
The lower panel shows the
integrated profile in total intensity (thick line), linear polarization
(dark grey line) and circular polarization (light grey line).
The choice of longitude zero for each pulsar at the symmetrical
midpoint of the profile and the profiles at the two frequencies 
have been aligned as closely as possible.}
\end{figure*}
\begin{figure*}
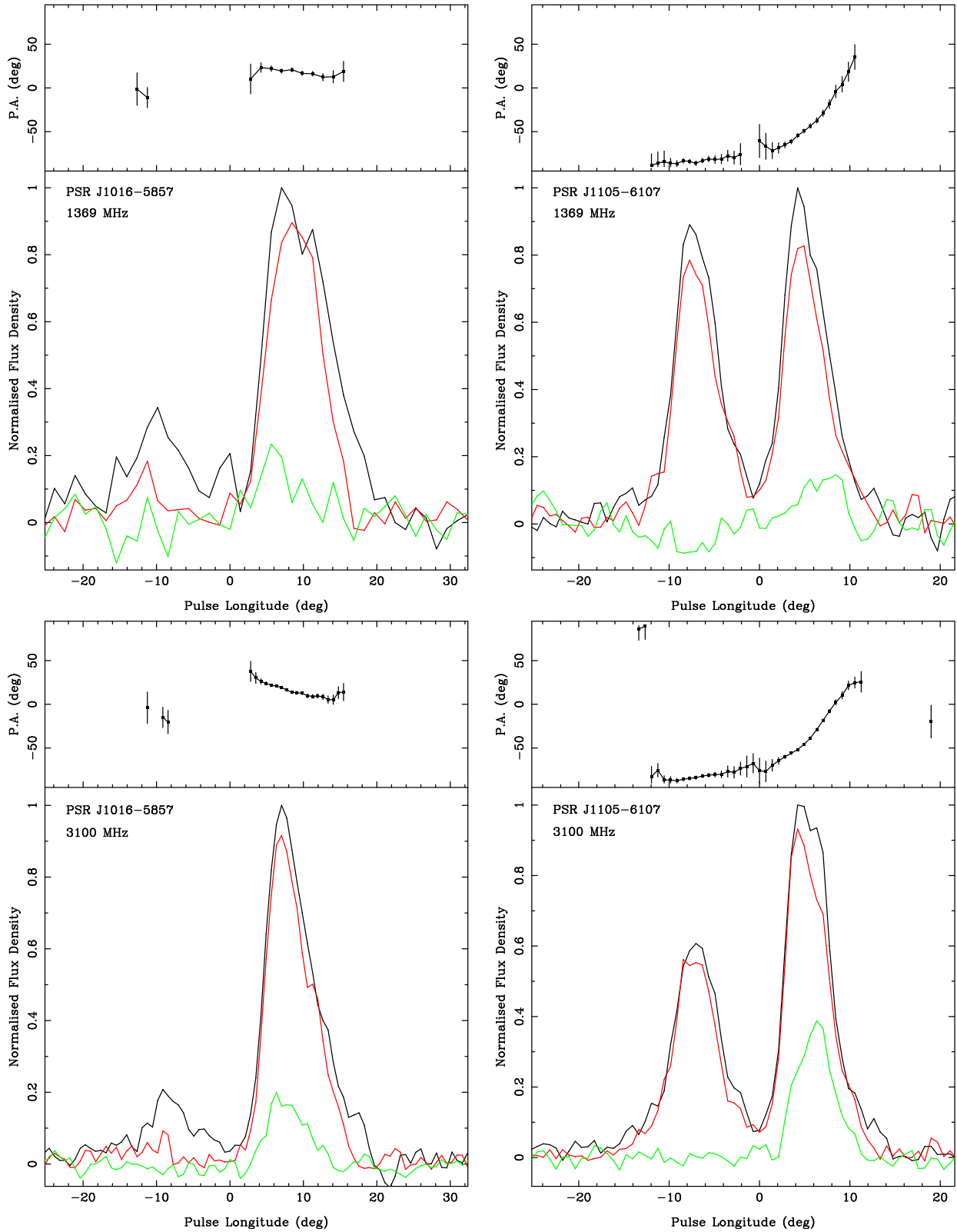

\begin{tabular}{cc}
\psfig{figure=fig2a.ps,angle=0,width=8cm} &
\psfig{figure=fig2b.ps,angle=0,width=8cm} \\
\psfig{figure=fig2c.ps,angle=0,width=8cm} &
\psfig{figure=fig2d.ps,angle=0,width=8cm} \\
\end{tabular}
\caption{Polarization profiles at 1.4~GHz and 3.1~GHz for
PSRs J1016$-$5857 and J1105$-$6107. See Fig~1 for details.}
\end{figure*}
\begin{figure*}
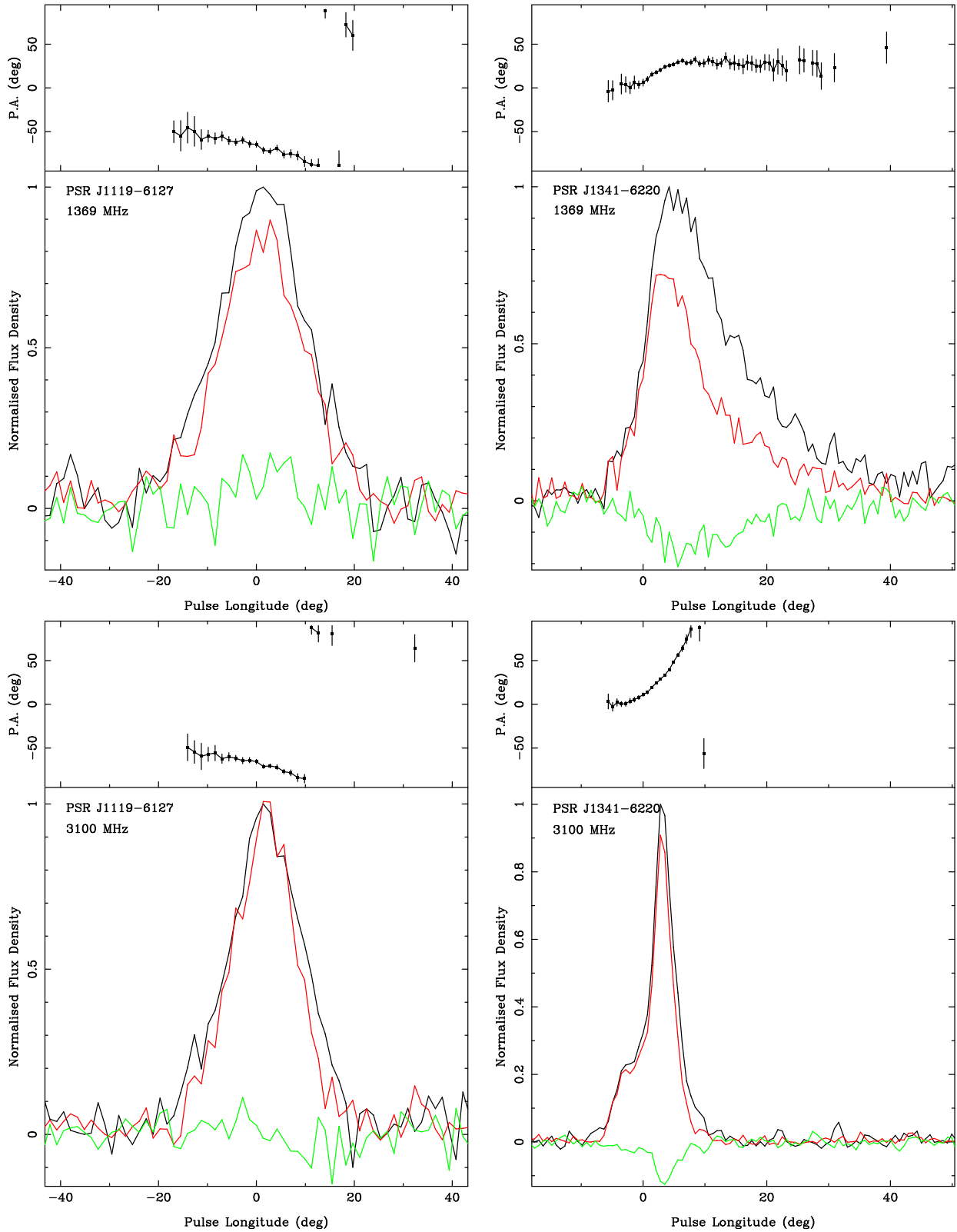

\begin{tabular}{cc}
\psfig{figure=fig3a.ps,angle=0,width=8cm} &
\psfig{figure=fig3b.ps,angle=0,width=8cm} \\
\psfig{figure=fig3c.ps,angle=0,width=8cm} &
\psfig{figure=fig3d.ps,angle=0,width=8cm} \\
\end{tabular}
\caption{Polarization profiles at 1.4~GHz and 3.1~GHz for
PSRs J1119$-$6127 and J1341$-$6220. See Fig~1 for details.}
\end{figure*}
\begin{figure*}
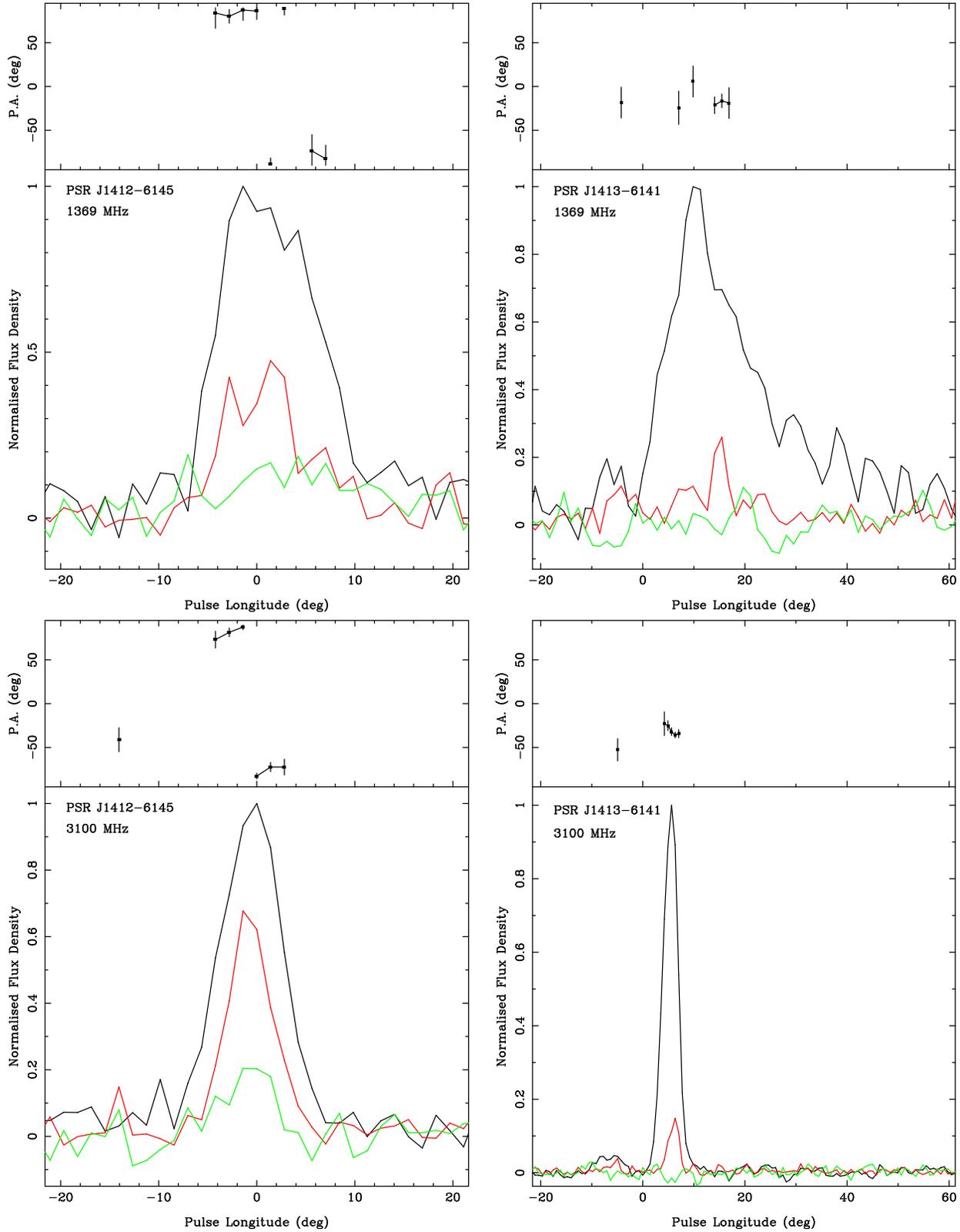

\begin{tabular}{cc}
\psfig{figure=fig4a.ps,angle=0,width=8cm} &
\psfig{figure=fig4b.ps,angle=0,width=8cm} \\
\psfig{figure=fig4c.ps,angle=0,width=8cm} &
\psfig{figure=fig4d.ps,angle=0,width=8cm} \\
\end{tabular}
\caption{Polarization profiles at 1.4~GHz and 3.1~GHz for
PSRs J1412$-$6145 and J1413$-$6141. See Fig~1 for details.}
\end{figure*}
\begin{figure*}
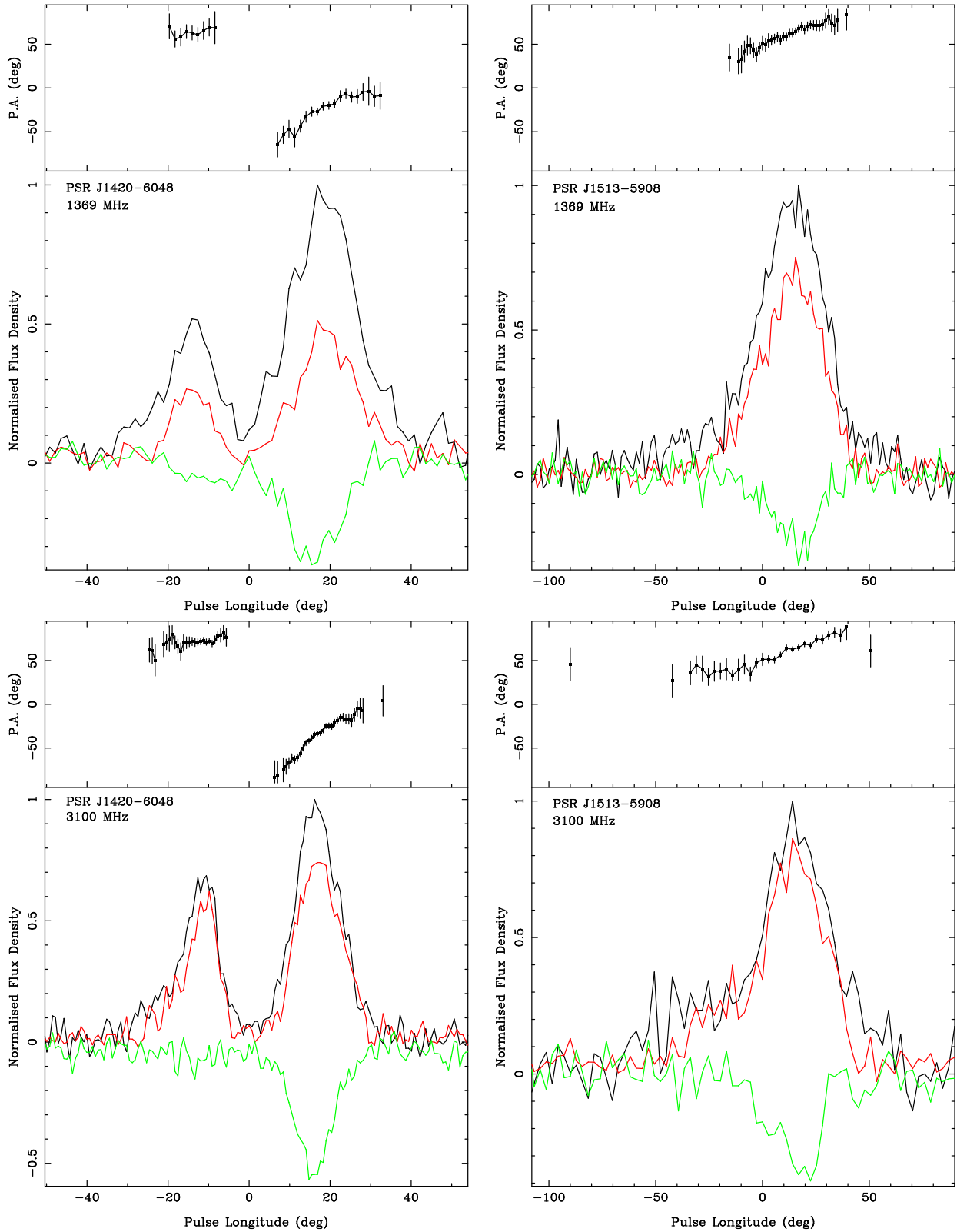

\begin{tabular}{cc}
\psfig{figure=fig5a.ps,angle=0,width=8cm} &
\psfig{figure=fig5b.ps,angle=0,width=8cm} \\
\psfig{figure=fig5c.ps,angle=0,width=8cm} &
\psfig{figure=fig5d.ps,angle=0,width=8cm} \\
\end{tabular}
\caption{Polarization profiles at 1.4~GHz and 3.1~GHz for
PSRs J1420$-$6048 and J1513$-$5908. See Fig~1 for details.}
\end{figure*}
\begin{figure*}
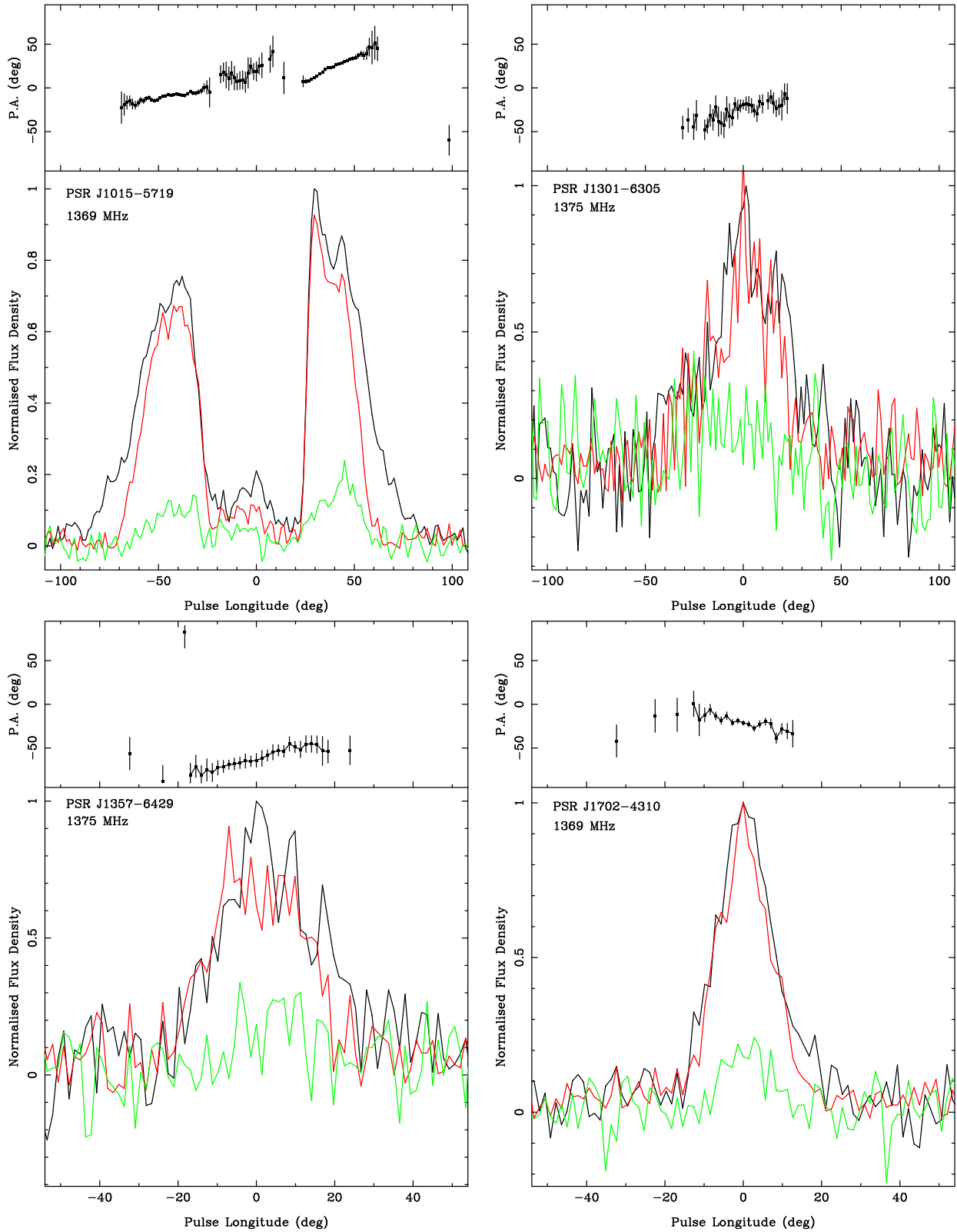

\begin{tabular}{cc}
\psfig{figure=fig6a.ps,angle=0,width=8cm} &
\psfig{figure=fig6b.ps,angle=0,width=8cm} \\
\psfig{figure=fig6c.ps,angle=0,width=8cm} &
\psfig{figure=fig6d.ps,angle=0,width=8cm} \\
\end{tabular}
\caption{Polarization profiles for PSRs J1015$-$5719, J1301$-$6305,
J1357$-$6429 amd J1702$-$4310 which were observed only at 1.4~GHz.
See Fig~1 for details.}
\end{figure*}

\noindent
{\bf PSR J1015$-$5719:} Observations were made only at 1.4~GHz for this
pulsar. The profile is a wide double with emission over nearly 200 degrees
of longitude.  A small central component is also visible.
It is interesting that the steep edges of
the profile appear on the interior rather than the exterior of the
components. This is the opposite to what is generally observed.
Circular polarization is present against both the
leading and trailing component. The linear polarization is similar against
both components; it is very high except for the wings of the pulse
where it is virtually zero.
This implies that both components must 
themselves be composed of several individual components.
We will return to this point later.
It appears likely that there is an orthogonal
jump between the central and trailing components.

\noindent
{\bf PSR J1016$-$5857:} Similar to PSR~J0940$-$5428, the profile shows
two components of which the trailing component is dominant.
The polarization is low in the leading component but relatively high
in the trailing component, as is the circular polarization.
The behaviour of the PA across the leading component is unclear.
Across the trailing component, the PA seems rather flat at 1.4~GHz
but appears to have significant negative slope at 3.1~GHz.
The profile is wider at the lower freqeuency.

\noindent
{\bf PSR J1105$-$6107:} The profile at 1.4~GHz is similar to that seen
in Crawford et al. (2001) except that our better time resolution clearly
splits the two components. The pulse profile consists of two
components, nearly equal in strength at 1.4~GHz but with the trailing
component dominant at 3.1~GHz. The linear polarization is high throughout
but circular polarization is only present under the trailing component.
The PA swing is flat under the leading component and steepens under 
the trailing component.

\noindent
{\bf PSR J1119$-$6127:} The pulse profile is a similar quasi-Gaussian
shape at both frequencies although the 3.1~GHz profile is slightly 
narrower than its 1.4~GHz counterpart. The percentage linear
polarization is high throughout the profie and the PA swing is flat.
\begin{table*}
\caption{Constraints on geometric angles and emission heights from RVM fits}
\begin{tabular}{llrrrrrrrrr}
\hline & \vspace{-3mm} \\
Jname & Bname & r$_{\rm lc}$ & $\alpha$ & $|\beta|$ & $\phi_0$ & PA$_0$ & h$_{\rm em}$ & $h_{\rm em}/r_{\rm lc}$ & $\rho_o$ \\
& & (km) & (deg) & (deg) & (deg) & (deg) & (km) & & (deg) \\
\hline & \vspace{-3mm} \\
J0729$-$1448 & & 12000 & -- & $<$9 & 10.3 & $-$82.2 & 630 & 0.053 & 20 \\
J0940$-$5048 & & 4200  & -- & $<$20 & 2.8 & $-$85.4 & 60 & 0.015 & 11 \\
J1015$-$5719 & & 6700  & 101$\pm$5 & 20$\pm$5 & 12.9 & 62 & 380 & 0.057 & 21 \\
J1105$-$6107 & & 3000  & -- & $<$4 & 8.5 & $-$2.2 & 110 & 0.037 & 16 \\
J1119$-$6127 & & 19500 & -- & $<$20 & 25.9 & -- & 2200 & 0.11 & 29 \\
J1341$-$6220 & B1338$-$62 & 9200 & -- & $<$5 & 6.9 & 72.5 & 280 & 0.031 & 15 \\
J1420$-$6048 & & 3300  & -- & $<$15 & 12.3 & $-$61.0 & 175 & 0.053 & 20 \\
J1513$-$5908 & B1509$-$58 & 7200 & -- & -- & 14.8 & 63.0 & 465 & 0.064 & 22\\
\hline & \vspace{-3mm} \\
\end{tabular}
\end{table*}

\noindent
{\bf PSR J1301$-$6305:} The profile consists of a single component
which occupies about 100 degrees of longitude. It is highly polarized
and shows a shallow swing of position angle across the pulse.
It most resembles the profile of PSR~J1513$-$5908 described in
more detail below. The pulsar was observed only at 1.4~GHz.

\noindent
{\bf PSR J1341$-$6220 (B1338$-$62):} The profile at 1.4~GHz is similar
to that published in Crawford et al. (2001). It is largely dominated
by the scattering tail (hence the large apparent width) though there is 
a hint of an initial leading component.
At 3.1~GHz the scattering is significantly reduced and the emerging
leading component can clearly be seen. The pulsar has very high
linear polarization and some circular polarization under the main
component. The PA swing is initially flat before rising steeply against
the trailing component.

\noindent
{\bf PSR J1357$-$6429:} The profile shows a single component with high
linear polarization and a shallow swing of position angle. There are
no observations at the higher frequency for this pulsar.

\noindent
{\bf PSR J1412$-$6145:}  The profile at 3.1~GHz is significantly narrower
than that at 1.4~GHz and the linear polarization is also higher at
the higher frequency. It seems likely that the trailing edge of the
lower frequency profile contains an unpolarized component which
is not present at the higher frequency.

\noindent
{\bf PSR J1413$-$6141:} The pulse profile at 1.4~GHz seems to consist
of a small leading component followed by a dominant second component
which is scatter broadened. At 3.1~GHz the higher signal to noise allows
the leading component to be seen more clearly. The dominant
component has some linear polarization but virtually no circular polarization
and a steep PA swing across it. The profile appears very narrow at
3.1~GHz, however, the initial leading component is below the 10 per cent
intensity level and would add another 15\degr\ to the width.
This is the only pulsar in the sample in which the linear polarization is low.
The derived value for the RM is necessarily uncertain.

\noindent
{\bf PSR J1420$-$6048:} The pulse profile at 1.4~GHz is similar to that
presented in Roberts et al (2001). There are two components, both
highly linearly polarized with significant circular polarization only
under the trailing component. The PA is flat under the leading component
but rises steeply under the trailing component. At 3.1~GHz the leading
component is slightly stronger relative to the trailing component than
at 1.4~GHz and overall the profile is narrower perhaps due to scattering
at the lower frequency.

\noindent
{\bf PSR J1513$-$5908 (B1509$-$58):}
The profile has a high degree
of linear polarization and a significant amount of circular polarization
at both 1.4 and 3.1~GHz.
In our data we see a hint of a leading component at 1.4~GHz which
has become readily apparent at 3.1~GHz.
The PA swing is initially flat before steepening under the trailing
component. The pulsar appears to be more highly linearly polarized
at 3.1~GHz than at 1.4~GHz. The width has also increased at 3.1~GHz
because of the emergence of the leading component.
The polarization profile at 1.4~GHz is similar to that in
Crawford et al. (2001) although we note that our polarization
fraction is somewhat lower than shown by those authors.

\noindent
{\bf PSR J1702$-$4310:} 
The pulsar has a single pulse component with a width of $\sim$35\degr.
It is highly linearly polarized and has significant circular
polarization. As is typical of these single-type pulsars, the PA swing
is unbroken and rather flat.

\section{Fitting the RVM}
We have attempted to fit the rotating vector model (RVM) to the 
position angle swing for each pulsar in our sample.
As is usual, $\alpha$
is poorly constrained in all cases (except for PSR~J1015$-$5719),
whereas a constraint can be placed on $\beta$ for about half the sample.
However, the location of $\phi_0$ can often be located with
good precision (at least in a statistical sense).
The results are the same within the errors at both 1.4 and 3.1~GHz for
those pulsars with dual frequency observations. The only exception to
this is PSR~J1341$-$6220 which is highly scattered at the lower
frequency and for which we did not attempt a fit.
Table~2 lists the pulsars from which meaningful
constraints were obtained from the fitting.
The third column gives the light cylinder radius of the pulsar,
columns 4 and 5 give the geometrical angles. Column 6 gives $\phi_0$, the
location of the inflexion point of the PA curve relative to the 
zero point of longitude taken from Figures~1-6.

\begin{figure*}
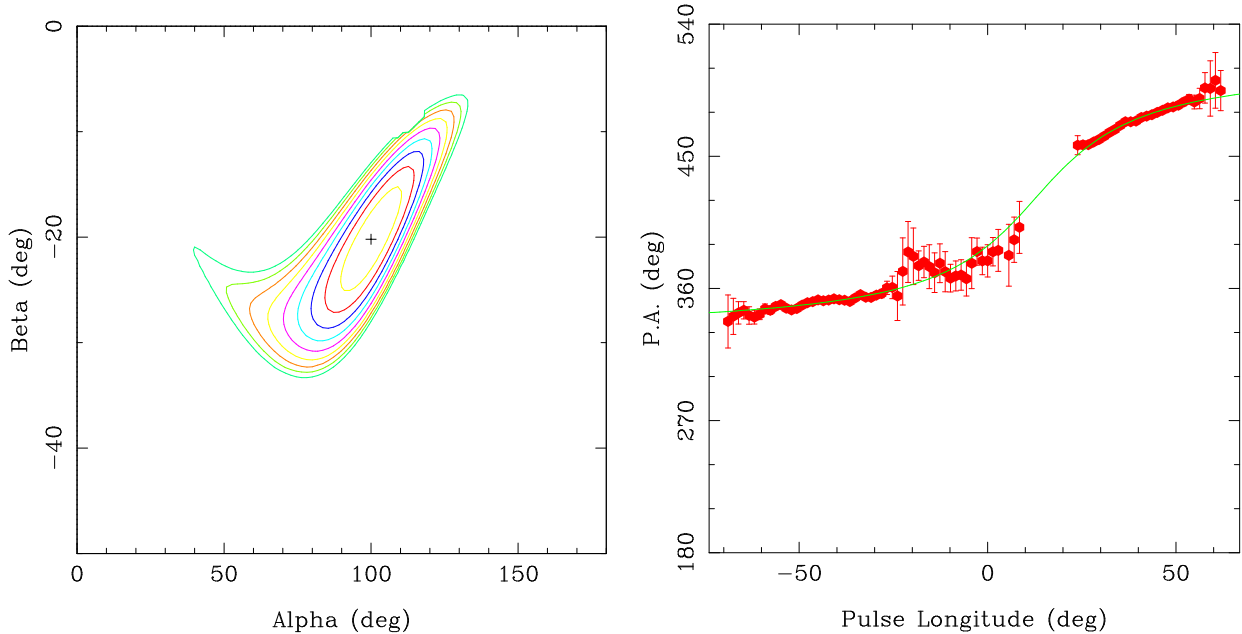

\begin{tabular}{cc}
\psfig{figure=fig7a.ps,angle=-90,width=8cm} &
\psfig{figure=fig7b.ps,angle=-90,width=8cm} \\
\end{tabular}
\caption{RVM fitting to PA swing in PSR~J1015$-$5719. The left hand panel shows
the $\chi^2$ contours in the $\alpha$-$\beta$ plane. The right hand panel
shows the data (with the addition of an orthogonal jump at longitude 
$\sim$15\degr, cf Fig.~6)
and the best fit with $\alpha$=101\degr, $\beta$=$-$20\degr.}
\label{rvmfit}
\end{figure*}

PSR~J1015$-$5719 is the only pulsar in the sample which provides
a good RVM fit with strong constraints on the geometric angles. This is because
the width of the profile is large and the 
presence of the central component means that
PA values can be obtained almost throughout the entire longitude range.
Even in this case, however, there are two possible RVM fits that can be made.
The best fit is obtained only after adding an orthogonal jump between the
central and trailing component. Figure~\ref{rvmfit} shows the $\chi^2$
contours in the $\alpha$-$\beta$ plane in the left hand panel, with
the right hand panel showing the data and the best fit to the PA
with $\alpha=101\degr, \beta=-20\degr$.
The addition of the orthogonal jump `forces'
$\phi_0$ to be $\sim$13\degr.
If the orthogonal jump is omitted, an RVM fit yields
$\alpha\sim 86\degr, \beta\sim -10\degr$ and the best fit line passes
underneath the majority of the central component PA values. In this case,
$\phi_0$ lies at a longitude of $\sim$250\degr, significantly after the
emission locations.
Whichever fit is chosen, small values of $\alpha$ are ruled out and
the pulsar appears to be an orthogonal rotator viewed at a rather 
large impact angle.
Note that the $\alpha$ and $\beta$ values quoted here are corrected for
the `PA convention problem' \cite{ew01}, wherein observed PAs increase
counter-clockwise while the RVM model assumes the opposite convention.

For PSR~J1119$-$6127, we get similar results to
Crawford \& Keim (2003)\nocite{ck03} in that $\beta$ must be less than about
20\degr, and $\alpha$ is almost unconstrained. As
they do, we also derive $\phi_0$ to
be on the far trailing edge of the pulse profile. Although there is no
linear polarization at this location we can estimate PA$_0$
to be $\sim$33\degr from the RVM fit.
For PSR~J1420$-$6048, we find that $\alpha$ and $\beta$ cannot be 
as well constrained as claimed by Roberts et al. (2001)\nocite{rrj01}
who derived $\alpha<35\degr$ and $\beta\sim0.5\degr$.
For PSR~J1513$-$5908, contrary to Crawford et al. (2001)\nocite{cmk01},
we cannot constrain $\alpha$, and we therefore cannot exclude the estimate
of 60\degr\ made by Romani \& Yadigaroglu (1995)\nocite{ry95} from the 
high energy pulse profile. Interpretation of the X-ray morphology of the 
pulsar wind nebula yields evidence that $\alpha+\beta$ is
greater then 70\degr \cite{bb97,ykk+05}. If this value is correct, our
fit then constrains $\alpha$ to lie between 30\degr\ and 150\degr.

\section{Polarization of young pulsars}
It is evident even from a cursory glance at Figs~1$-$6 that many of the
profiles of these young pulsars have remarkably similar characteristics.
First, we see
that these pulse profiles are relatively simple, unlike the highly complex
profiles seen in older pulsars (as already pointed out decades ago
by Huguenin et al. 1971\nocite{hmt71}). Lyne \& Manchester (1988)
concluded that this was because the relative spectral index of core
and cone component was age dependent.
Of the 14 pulsars
studied, 9 are double profiled with the trailing component dominating
in all cases. Some have almost equal amplitude components
(e.g. PSR~J1420$-$6048) but in others the ratio
is rather large (e.g. PSR~J1513$-$5908).
Seven of the 9 which have measurable circular polarization have so 
only under the trailing component.
All 9 show a flat PA swing
followed by a steeper swing under the trailing component.
The five pulsars which do not have multi-component profiles are
PSRs J1119$-$6127, J1301$-$6305, J1357$-$6492, J1412$-$6145
and J1702$-$4310.
These have single components
and a rather flat PA swing with little or no circular polarization.
Finally, 12 of the 14 have large degrees of linear polarization (in excess of
50 per cent) and no obvious orthogonal mode jumps in the profiles
with the exception being PSR~J1015$-$5719 and possibly PSR~J1413$-$6141.
The spectral index evolution between these two frequencies shows no
obvious trends in the 8 pulsars with two components and dual frequency
measurements. Three show a flatter
index leading component, three have a flatter index trailing component
and the remaining two have similar indicies for both leading and
trailing components!

We note that these features are not only seen in the present data
but also in previous polarization observations of young pulsars
\cite{qmlg95,hkk98,cmk01,kjm05}.
Nearly all have high linear polarization and those pulsars with
two components have the trailing component dominant (e.g. PSRs
B1610$-$50, B1643$-$43, B1758$-$23, B1800$-$21, B1823$-$13). The only
exception to this rule is the Vela pulsar in which the leading component
dominates the profile below 5~GHz.

There are two possible explanations for the morphology of these young pulsars.
In the first scenario, the two observed components represent
the leading edge of a 
cone and a more central component, with the trailing edge entirely absent.
In this case, one expects an asymmetric PA swing (as seen) and also increased
circular polarization under the central component (as also seen).
Generally, however,
conal components have a flatter spectral index than core components
and under this picture one might expect the leading component to 
dominate at higher frequencies (since conal components tend to dominate
over core components at high frequencies) and this is not obviously the case.
However, observations over a wider frequency range (especially very
low frequencies) might help solve this puzzle.
One also has to explain why
the trailing edge is entirely absent in these pulse profiles, but
this follows the trend noted by Lyne \& Manchester (1988)\nocite{lm88}
that leading edge cones dominate over trailing edge.

The second possibility is that the two principal observed components
mark `classical' double emission
with the magnetic pole crossing near the symmetrical centre of
the pulse profile. In this case, the asymmetry of the PA swing is then
due to A/R which has a greater effect in rapidly spinning pulsars
than in slowly spinning ones. The question arises as to why the trailing
component always dominates and has significant circular polarization.
The single component pulsars would then either be an extreme case where the
leading component is absent and only the trailing component is present,
or a grazing cut through the outer edge of the emission cone. The
latter seems more plausible in light of the flat PA swings seen
in the single component pulsars.
This explanation goes against the trend shown by
Lyne \& Manchester (1988)\nocite{lm88} where, in pulsar with partial cones,
generally the leading component dominates.
Taking all the data on young pulsars into account, 
we also violate the correlation (in 9 out of 15 cases)
between the handedness of circular polarization
and the sense of PA swing in conal pulsars found by Han et al. (1998).
On balance however, we favour this possibility for a number of reasons.
First that PSR~J1015$-$5719 clearly shows a component between the two
main components. Secondly, both components generally have the same width,
and the symmetry point lies between the components. Finally, conal
components generally have flat spectral indices as do young pulsars
in general.

\subsection{Emission Heights}
If we assume the latter interpretation is the correct one, then we can
use the results of the RVM fitting to derive the emission height of
the radiation. We do this by first deriving the longitude of the
midpoint of the pulsar profile using its inherent symmetry, which
we denote $\phi_s$. Note that in Figures 1$-$6 we have located $\phi_s$
at zero longitude as far as possible.
Then, we use the value of $\phi_0$ given in Table~2 to obtain the emission
height by computing $\delta\phi({\rm PA})= \phi_0 - \phi_s$
and attributing the difference to the effects of A/R. 
Application of equation~2 then leads to a determination of the emission
height. These heights are listed in column~8 of Table~2; column~9 lists
the heights in terms of the fraction of the light cylinder radius.
From the emission heights we can compute the cone half-opening angle
at the last open field line, $\rho_o$,
via equation 4. These values are shown in column 10 of Table~2.
The values of $\rho_o$ are distributed between 10 and 30 degrees.

The results show that the emission height for all these pulsars
lies between 5 and 15 per cent of the light cylinder. This is significantly
higher than the 1-2 per cent heights found for older pulsars \cite{gg03}.
We caution however against interpreting these values
too literally. In recent years evidence has built up that the emission
height at a given frequency is not constant across
the beam but is higher at the edges
of the cone and lower in the middle \cite{gg03}. If this is the case,
A/R affects the core and cone by different amounts and
the symmetrical midpoint of the profile will not necessarily
correspond to the core location and the emission heights computed
through A/R will be underestimates. Finally, other effects
such as refraction effects \cite{pet00} and magnetic field sweepback can 
distort this picture still further \cite{ae98,dh04}.

We note there is a strong inconsistency in the
interpretation of the data on PSR~J1015$-$5719. The RVM fit shows the
pulsar is nearly an orthogonal rotator; if this is the case then
the very large pulse width 
naturally implies the emission must come from high in the magnetosphere.
On the other hand, the value of the emission height derived from the
location of $\phi_0$ is rather small.
There are two possible sources of error or confusion. The RVM fit
might be incorrect and the pulsar is really an aligned, rather than
orthogonal rotator. This is possible, witness for example the continuing
discussion on whether PSR~B0950+08 is aligned or orthogonal \cite{ew01},
but seems unlikely given the goodness of fit.
Secondly, assigning the offset in $\phi_0$ as arising due to A/R
is in error due to the reasons we suggested above.

However, there exists a further appealing solution for PSR~J1015$-$5719.
This comes from recognising that the measured A/R yields
the emission height 
{\it of the core emission} and may not reflect the emission height
of the conal emission if this occurs significantly higher in the
magnetosphere than the core emission \cite{gg03}. To test this,
we have therefore decomposed the profile of PSR~J1015$-$5719 into
7 Gaussians, 3 for the leading edge, a central component and 3 for
the trailing edge. The location of the centroid of these Gaussians
occurs at longitudes $-$72\degr, $-$52\degr, $-$36\degr, 0\degr,
31\degr, 44\degr\ and 56\degr (cf Figure~6). Immediately one can see that the
locations are asymmetric, with the trailing components closer to the
core component than the leading components, exactly as expected in
the picture of Gangadhara \& Gupta (2001)\nocite{gg01}.
We can use this asymmetry, in conjunction with the derived emission
height of $\sim$400~km for the core component to derive the heights
of the 3 conal components (see equation~5).
These are (from inner to outer) 550~km,
650~km and 850~km respectively. Using the more exact formalism of
Gangadhara (2005) which takes into account the geometrical angles the
height of the outermost cone increases to 925~km.
This helps solves the problem of
having an orthogonal rotator with a seemingly small emission height
and a wide profile. Indeed the emission height of the core is small,
but the outer cones are at a significantly higher height, yielding
a wide profile.
\begin{figure}
\psfig{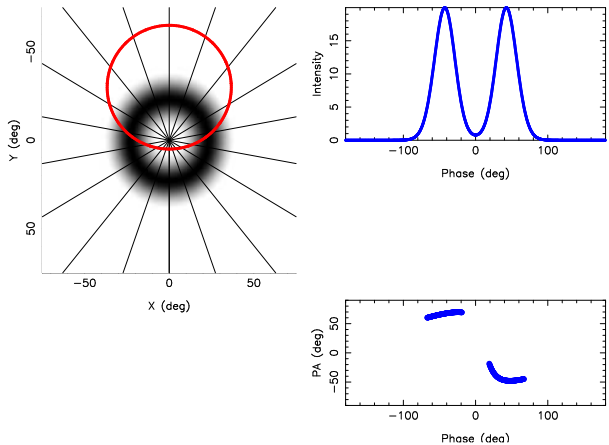}
\caption{Simulation of a pulse profile and PA swing from our simple model.
Parameters are $\alpha$=30\degr, $\beta$=5\degr, and we assume a pulse
period of 100~ms with an emission height of 250~km. The left panel shows
the emission pattern with the line of sight traverse superposed.
The top right panel shows the resultant pulse profile with the bottom
right panel showing the PA swing, which is delayed relative
to the total intensity profile by an amount consistent with
the emission height and the effects of A/R.}
\label{beam}
\end{figure}

\subsection{A Generic Young Pulsar Beam Model}
We have concluded that young pulsar beams generally consist of a hollow cone,
with little or no emission from the center. The double profiles
arise from cuts close to the magnetic pole, the single profiles
are grazing trajectories along the cone. We quantify this by adopting
a canonical young pulsar with a spin period of 0.1~s. The light cylinder
radius is at $\sim$5000~km with the emission zone located at $\sim$250~km,
consistent with the results in Table~2. These parameters naturally
lead to a beam half-opening angle of $\sim$20\degr.
A simple beam model can then be created in which
$\rho$=20\degr\ and there is a single hollow cone
of emission located at the beam edge. The conal
emission zone is relatively wide, has
a Gaussian amplitude distribution and
takes up $\sim$0.2 of the beam. There is no core emission. We then choose
random values of $\alpha$ and $\beta$ and compute the line of sight relative
to the beam in order to produce a simulated pulse profile.
At the same time, we can construct a PA swing based on the geometric angles,
assuming the RVM is correct. The PA swing is then delayed in longitude
with respect to the profile by an amount consistent with the emission
heights and the effects of A/R.
Fig~\ref{beam} shows an example. We find from the simulations that
$\sim$50 per cent of pulsars are potentially detectable
(i.e. we have some sightline across the open field lines); of these $\sim$45 per
cent have double profiles and $\sim$55 per cent have single profiles.
Most of the double component pulsars have widths of $\sim$40\degr, 
with a small minority having much wider beams. For the single components,
there is a ratio of 2:1 between narrow and wide profiles.
The results from this simple simulation are very similar to the observed
results. In the observed sample, there is roughly an equal split between
profiles with single components and those with two components.
Most of the double component profiles have similar widths, with the
occasional wide double. 
In contrast the single component pulsars show a larger variety in
pulse widths as seen in the simulation.

\section{Derived proper motion directions and their implications}
As discussed in the introduction, we can potentially determine
the direction of motion of the pulsars through knowledge of the
position angle of the rotation axis under the assumption that
they are aligned with the velocity direction. Johnston et al. (2005)
pointed out that, even if the vectors are aligned,
the polarization PA at $\phi_0$ could be
perpendicular to the rotation axis if the pulsar emits in the
orthogonal mode. We have seen from 
the discussion of the pulse profiles,
it is very difficult to correctly
assess the longitude of the magnetic pole in these young pulsars.
We can only really be confident in the cases where the RVM fit returns
an accurate value of $\phi$ and hence PA$_0$ as listed in Table~2 above.
This is only the case for 7 pulsars in our sample, and one of these
(PSR~J1119$-$6127) has no polarization at the location of $\phi_0$.
We describe the other six pulsars here.
The value of PA$_0$ has an inherent 180\degr\ ambiguity because of
the definition of position angle of linear polarization. Also, it is
possible that PA$_0$ can be 90\degr\ different from the true value of
the rotation axis position angle, if the pulsar is emitting in the
orthogonal mode (for full details see Johnston et al. 2005\nocite{jhv+05}).
Hence in what follows, we assume that either PA$_0$ or PA$_0\pm 90\degr$
will be parallel to the velocity vector of the pulsar.
This implicitly assumes that velocity
vector is aligned with the rotation axis as found by Johnston et al. (2005)
but we caution that this might not be a general rule.
\begin{figure}
\psfig{figure=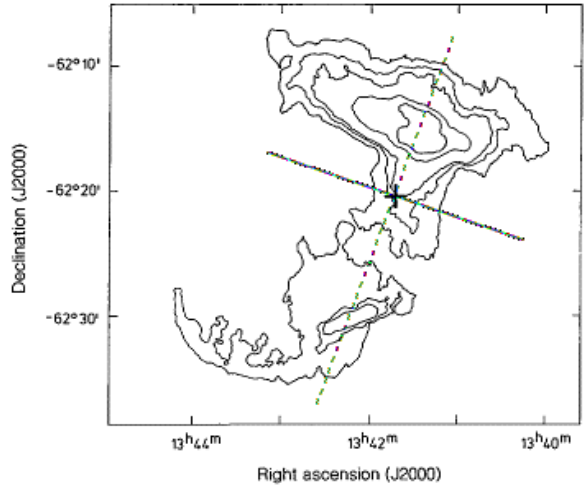,angle=0,width=8cm}
\caption{Radio continuum image of SNR G308.8$-$0.1 (Caswell et al. 1992).
The plus marks the position of PSR~J1341$-$6220. The solid line denotes
PA$_0$ from our measurements. The dashed line is orthogonal to 
PA$_0$ and is our preferred proper motion direction.}
\label{g308}
\end{figure}

\noindent
{\bf PSRs J0729$-$1448 and J0940$-$5048:} These two pulsars have no
apparent associations with high energy emission or SNRs. The value
of PA$_0$ and hence the direction of the rotation axis is parallel
or perpendicular to $-$82.2 and $-$85.4 degrees respectively.

\noindent
{\bf PSR J1105$-$6107:} In this pulsar, we showed in Section 2 that
one might expect the velocity vector to point at 135\degr\ if the
pulsar was associated with SNR~G290.1$-$0.8. The value of PA$_0$ that
we derive from the RVM fit, $-$2.2\degr, would then seem to imply that
the pulsar did not originate from SNR~G290.1$-$0.8. The pulsar has a 
relatively large characteristic age and it may be that its parent SNR
has simply dissipated.
\begin{figure}
\psfig{figure=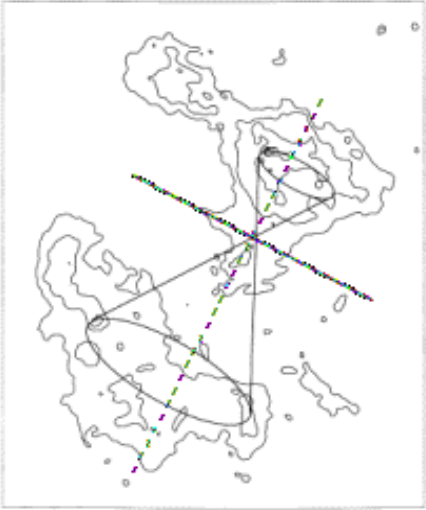,angle=0,width=6cm}
\caption{Radio continuum image of SNR G320.4$-$1.2.
PSR~J1513$-$5908 lies at the intersection of the two cones 
which represent a possible outflow geometry based on the continuum
morphology (Gaensler et al. 1999). The solid line denotes
PA$_0$ from our measurements. The dashed line is orthogonal to 
PA$_0$ and is our preferred proper motion direction.}
\label{g320}
\end{figure}

\noindent
{\bf PSR J1341$-$6220:} The main axis of symmetry of the peculiar SNR
in which PSR~J1341$-$6220 is embedded has a position angle of $-$20\degr,
which is $\sim$90\degr\ offset from our measured PA$_0$ of 72.5\degr.
It seems likely therefore that the pulsar is moving along this symmetry axis,
the rotation axis also points in this direction and it emits in
the orthogonal mode.
Figure~\ref{g308} shows the SNR/PSR association with our derived 
value of PA$_0$ and the preferred proper motion direction.

\noindent
{\bf PSR J1420$-$6048:} The pulsar lies in a highly complex region of
the Galactic plane as detailed above. A proper motion in virtually any 
direction would intersect something of interest! It so happens that
the perpendicular to our PA$_0$ of $-$61\degr\ 
points at the centre of the `Kookaburra' complex.
Is PSR~J1420$-$6048 a pulsar moving at $\sim$400~km~s$^{-1}$ which
originated in the centre of the `Kookaburra', punctured through the
shell and is currently creating a wind bubble similar to the picture
seen in the `Duck' system \cite{gf00,tbk02}?

\noindent
{\bf PSR J1513$-$5908:} The PA of the symmetry axis of the PWN in
which the pulsar is embedded is 150\degr$\pm$5\degr. Gaensler et al. (2002)
\nocite{gak+02} infer that the rotation axis of the pulsar must also
have this PA to conform with their model. Our polarization measurements
yield PA$_0$ of 63\degr, which differs by 87\degr\ from the Gaensler et al.
(2002) conjecture. We therefore support the idea that the PWN symmetry
axis is indeed aligned with the pulsar's rotation axis and that the pulsar
emits in the orthogonal mode as in the Vela pulsar (Johnston et al. 2005)
and the above case of PSR~J1341$-$6220. If the correlation between
the spin axis and the velocity vector in pulsars is also correct then
we predict the pulsar's proper motion will also lie along this axis.
Figure~\ref{g320} shows the SNR/PSR association with 
our derived value of PA$_0$ and our proposed proper motion direction.

\section{Conclusions}
We have observed 14 young pulsars and obtained calibrated polarization
profiles. The profiles of young pulsars are generally simple and
fall into two categories. In the first, the profiles consist of two,
highly polarized components with a flat PA swing across the first component
and a steep swing across the second.
The second component is nearly always brighter than the first component.
The correlation pointed out by Han et al. (1998) between the sign of
circular polarization and the sense of swing of the PA appears not
to apply in these double pulsars.
The second category consists of
a simple, broad single component with high polarization and a shallow
swing of position angle.

We interpret these results as showing that the emission beams of young
pulsars consist simply of a single emission cone located near the last
open field lines. Core emission is absent or rather weak.
Emission arises between 1 and 10 percent of the light cylinder radius
which is significantly higher than the emission zone in older pulsars.
This simple model explains many of the features of the observations.
Two questions remain: why are the trailing components always brighter and
why do they generally have more circular polarization?

We predict that PSRs J1341$-$6220 and J1513$-$5908 have aligned rotation
and velocity axes and it would be useful to measure the
proper motion of these pulsars as verification.

\section*{Acknowledgments}
The Australia Telescope is funded by the Commonwealth of 
Australia for operation as a National Facility managed by the CSIRO.
JMW was supported by Grant AST 0406832 from the  U.S. National Science 
Foundation.  We thank M.~Kramer for the RVM fitting routines.
We thank the referee, R.~T.~Gangadhara, for his careful reading of
the manuscript.

\bibliography{modrefs,psrrefs,crossrefs}
\bibliographystyle{mn}
\label{lastpage}
\end{document}